\documentclass[aps,prl,twocolumn,showpacs,floatfix,preprintnumbers,amsmath,amssymb,nofootinbib,superscriptaddress]{revtex4-1}

\input epsf
\usepackage{graphicx}
\usepackage{color}
\usepackage{mathtools}

    \newcommand{\be}{\begin{equation}}
  \newcommand{\ee}{\end{equation}}
    \newcommand{\ba}{\begin{align}}
  \newcommand{\ea}{\end{align}}

\newcommand{\Msun}{M_{\odot}}
\newcommand{\fDM}{f_{\rm DM}}

\begin{document}

\title{Lensing of Fast Radio Bursts as a Probe of Compact Dark Matter}

\author{Julian B. Mu\~noz}
\author{Ely D. Kovetz}
\affiliation{Department of Physics \& Astronomy, Johns Hopkins University, 3400 N. Charles 
	St., Baltimore, MD 21218, USA}
\author{Liang Dai}
\affiliation{Institute for Advanced Study, Einstein Drive, Princeton, New Jersey 08540}
\author{Marc Kamionkowski}
\affiliation{Department of Physics \& Astronomy, Johns Hopkins University, 3400 N. Charles 
	St., Baltimore, MD 21218, USA}

\date{\today}

\begin{abstract}	
The possibility that part of the dark matter is made of massive compact halo objects 
(MACHOs) remains poorly constrained over a wide range of masses, and especially 
in the $20-100\, M_\odot$ window. We show that strong gravitational lensing of 
extragalactic fast radio bursts (FRBs) by MACHOs of masses larger than 
$\sim20\,M_\odot$ would result in repeated FRBs with an observable time delay. 
Strong lensing of a FRB by a lens of mass $M_L$ induces two images, separated 
by a typical time delay $\sim$ few $\times(M_L/30\, M_\odot)$ milliseconds.
Considering the expected FRB detection rate by upcoming experiments, such as CHIME,
of $10^4$ FRBs per year,
we should observe from tens to hundreds of repeated bursts yearly, if MACHOs in this window make up
all the dark matter. 
A null search for echoes with just $10^4$ FRBs would constrain the fraction $f_{\rm DM}$ of dark matter in MACHOs to $f_{\rm DM}\lesssim 0.08$ for $M_L\gtrsim 20\,M_\odot$.

\end{abstract}

\maketitle

Although observations indicate that dark matter accounts for a significant 
share of the energy density of our Universe \cite{1502.01589}, we do not 
know its composition. 
Longtime candidates to make up the dark matter are massive compact halo 
objects (MACHOs) \cite{astro-ph/0011506}.
They were originally proposed to be as light as $10^{-7}\Msun$ and as heavy as 
the first stars ($\sim\,10^3\Msun$) \cite{Griest}.
Over the years, different experiments have progressively constrained the
fraction $\fDM$ of dark matter that can reside in MACHOs with a given mass, 
placing tight upper bounds over most of the vast range above. 
High-mass ($\gtrsim 100\, \Msun$) MACHOs, for example, are constrained by 
the fact that they would perturb wide stellar binaries in our Galaxy \cite{0903.1644}.
Meanwhile, lower-mass ($\lesssim 20 \Msun$) MACHOs are 
effectively ruled out as the sole component of Galactic dark matter, as they would 
create artificial variability in stars, due to gravitational microlensing 
\cite{astro-ph/0607207,1106.2925,1504.05966}.

However, there remains a window of masses between 
$20$ and $100\,\Msun$, where the constraints are weaker, and 
in which arguably all the cosmological dark matter could be in the 
form of MACHOs \cite{Pooley:2008vu,0910.3645,1106.2925,1406.5169,1504.05966}.
This is a particularly interesting window, as it has been recently argued in  
Ref.~\cite{1603.00464} that if primordial black holes (PBHs) \cite{Hawking,Carr} 
in the $\sim30\,\Msun$ mass range are the constituents of dark matter, 
they form binaries in halos, coalesce, and emit observable gravitational waves, 
with an event rate consistent with the published LIGO detection \cite{LIGO}. 

In this Letter we propose to use the strong lensing of fast radio bursts 
(FRBs) to probe MACHOs of masses $\gtrsim 20 \,\Msun$, including PBHs, 
and either confirm that they make up the dark matter or close this window. 
FRBs are strong radio bursts with a very short duration, 
which makes them ideal as microlensing targets. Their temporal width is increased 
by the dispersion measure (DM), which measures the time 
delay of photons with different radio frequencies due to scattering by free electrons 
on their way to Earth. All detected FRBs to date possess high DMs, which yield 
burst widths of $\sim 1-10$ ms \cite{Lorimer:2007qn,Keane:2011mj,1307.1628,
1404.2934,1412.0342,1412.1599,1511.07746,Thornton(2013),1601.03547,1603.00581}. 
These values of the DM are several times larger than the expected contribution from 
free electrons
within the Milky Way \cite{astro-ph/0207156,1405.5945}, suggesting their origin
is extragalactic (some authors, however, prefer a Galactic origin \cite{1310.2419,1507.01002}). 
Proposed sources of extragalactic FRBs include merging neutron stars \cite{astro-ph/0003218} 
or white dwarfs \cite{1307.7708}, as well as bursts from pulsars \cite{1501.00753}.

Strong lensing of a FRB by a MACHO will generate two images of the burst. 
While their angular separation may be too small to be resolved, the time delay between
them, on the order of milliseconds for a MACHO lens with mass $M_L\sim 20-100\,\Msun$, 
might be large enough to enable a detection of two separate peaks, rather than one, if the time delay is bigger than the pulse width. 
Fortunately, the lensing of FRBs by compact objects is not necessarily an unlikely occurrence. 
In fact, if all the dark matter is in MACHOs, roughly one in 50 FRBs originating at $z=0.5$ should be lensed.  If there are $\sim10^4$ FRBs on the full sky each day \cite{1602.07292}, then as many as $\sim 20$ microlensed FRBs may be reaching Earth daily.  Upcoming surveys, like APERTIF \cite{0806.0234}, UTMOST \cite{1601.02444}, HIRAX \cite{Newburgh:2016mwi}, or CHIME \cite{1406.2288}, which will map a considerable fraction of the sky, may thus see a significant number of lensed FRBs.

Below, we calculate the effects of microlensing on a given FRB and compute the optical
depth for strong lensing by compact objects. We then combine those results with different 
redshift distributions of FRBs 
and estimate how many lensed bursts are expected if MACHOs 
make up all the dark matter.  We also estimate the smallest 
fraction $f_{\rm DM}$ that will give rise to a detectable 
rate of microlensed events.

A MACHO of mass $M_L$ can be treated as a point lens with an (angular) Einstein radius:
\be
\theta_E = 2 \sqrt{ \dfrac{G M_L}{c^2} \dfrac{D_{LS}}{D_S D_L}},
\ee
where  $D_S$, $D_L$, and $D_{LS}$ are the (angular-diameter) distances to the source, to the lens, 
and between the source and the lens, respectively \cite{astro-ph/9606001,astro-ph/0305055}. A point 
lens produces two images, at positions 
$\theta_{\pm} = (\beta \pm \sqrt{\beta^2 + 4 \theta_E^2})/2$, 
where $\beta$ is the (angular) impact parameter. The time delay between these two images 
is
\be
\Delta t = \! \dfrac{4 \,G M_L}{c^3} (1+z_L)\!  \left [ \dfrac y 2 \sqrt{y^2+4} + \log\left(\dfrac{\sqrt{y^2+4}+y}{\sqrt{y^2+4}-y}\right)\right],
\label{eq:timedelay}
\ee
where $y\equiv \beta/\theta_E$ is the normalized impact parameter and $z_L$ is the redshift
of the lens. We also define the flux ratio $R_f$ as the absolute value of the ratio of the 
magnifications $\mu_+$ and $\mu_-$ of both images; i.e.,
\be
R_f \equiv  \left |\dfrac{\mu_+}{\mu_-} \right| = \dfrac{y^2+2+y\sqrt{y^2+4}}{y^2+2-y\sqrt{y^2+4}}>1.
\label{eq:Rf}
\ee

To claim that a FRB is strongly lensed we will require three conditions. First is that the brighter image
has a signal-to-noise ratio of 10 \cite{1412.0342}. Second is that the observed 
time delay is larger than some reference time $\overline {\Delta t}$, which will place a lower 
bound on the impact parameter $y>y_{\rm min} (M_L,z_L)$, calculated via Eq.~\eqref{eq:timedelay}. 
Finally, we demand that the flux ratio $R_f$ is smaller than some critical $\overline R_f$ (which we
take to be redshift independent), to ensure that both events are observed (note that for the fainter 
image the look-elsewhere effect is no longer relevant).
This forces the impact parameter to be 
smaller than $y_{\rm max} = \left [(1+\overline R_f)/\sqrt{\overline R_f}-2\right]^{1/2}$.

We now calculate the probability for a FRB to be lensed.
The lensing optical depth
of a source at redshift $z_S$ is given by
\be
\tau(M_L,z_S) = \int_0^{z_S} d\chi(z_L) (1+z_L)^2 n_L\, \sigma (M_L,z_L),
\label{eq:tauint}
\ee
where $\chi(z)$ is the comoving distance at redshift $z$, $n_L$ is the comoving number 
density of lenses, and $\sigma$ is the lensing cross section of a point lens of mass $M_L$, 
given by an annulus between the maximum and minimum impact parameters by
\be
\sigma (M_L,z_L) = \dfrac{4 \pi G M_L}{c^2} \dfrac{D_L D_{LS}}{D_S} \left[ y_{\rm max}^2 - 
y_{\rm min}^2(M_L,z_L)\right].
\ee
Equation~\eqref{eq:tauint} can be recast by using the Hubble parameter both at the redshift of the 
lens, $H(z_L)$, and today, $H_0$, as
\ba
\tau (M_L,z_S) &= \dfrac 3 2 \fDM \Omega_c \int_0^{z_S} dz_L \dfrac{H_0^2}{c \, H(z_L)} 
\dfrac{D_L D_{LS}}{D_S} \nonumber \\
 & \times (1+z_L)^2\left[ y_{\rm max}^2 - y_{\rm min}^2(M_L,z_L)\right],
\label{eq:tau}
\end{align}
where $\Omega_c=0.24$ is the cold-dark-matter density today, and the only remaining 
dependence on the lens mass $M_L$ is through $y_{\rm min}$.
Lower MACHO masses result in a lower optical depth, 
especially at lower source redshifts, due to our minimum time-delay requirement.

To calculate the integrated lensing probability, the optical depth for lensing of a single 
burst has to be convolved with the redshift distribution of incoming FRBs. We will 
consider two possible redshift distributions. 
First, we assume FRBs have a constant comoving number density, in which case the 
number of FRBs in a shell of width $dz$ at redshift $z$ is proportional to the 
shell's comoving volume $dV(z)=\left [4\pi\chi^2(z)/H(z) \right]dz$ \cite{1604.03909}, divided by $(1+z)$ 
to account for the effect of cosmological time dilation in the rate of bursts. 
To represent an instrumental signal-to-noise threshold we introduce a Gaussian cutoff 
at some redshift $z_{\rm cut}$, so the constant-density redshift distribution function 
would be
\be
N_{\rm const}(z) = {\cal N_{\rm const}} \dfrac{\chi^2(z)}{H(z)(1+z)} e^{-d_L^2(z)/[2 d_L^2(z_{\rm cut})]},
\label{eq:N(z)}
\ee
where $d_L$ is the luminosity distance, and $\cal N_{\rm const}$ is a normalization factor to ensure 
that $N_{\rm const}(z)$ integrates to unity.
Second, we consider a scenario in which FRBs follow the star-formation history (SFH) 
\cite{1512.02738}, whose density is parametrized as 
\be
\dot \rho_* (z)= h \dfrac{a+b z}{1+\left(\frac{z}{c}\right)^d},
\ee
with $a=0.0170$, $b=0.13$, $c=3.3$, $d=5.3$, and $h=0.7$ \cite{astro-ph/0012429,
astro-ph/0601463}. In this case, the SFH-based redshift distribution function $N_{\rm SFH}(z)$ is, 
\be
N_{\rm SFH}(z)={\cal N_{\rm SFH}} \dfrac{\dot \rho_* (z) \,\chi^2(z)}{H(z)(1+z)} e^{-d_L^2(z)/[2 d_L^2(z_{\rm cut})]},
\ee
and the normalization factor $\cal N_{\rm SFH}$ is chosen to have $N_{\rm SFH}$ integrate to unity. 

\begin{figure}[htbp!]
	\includegraphics[width=0.43\textwidth]{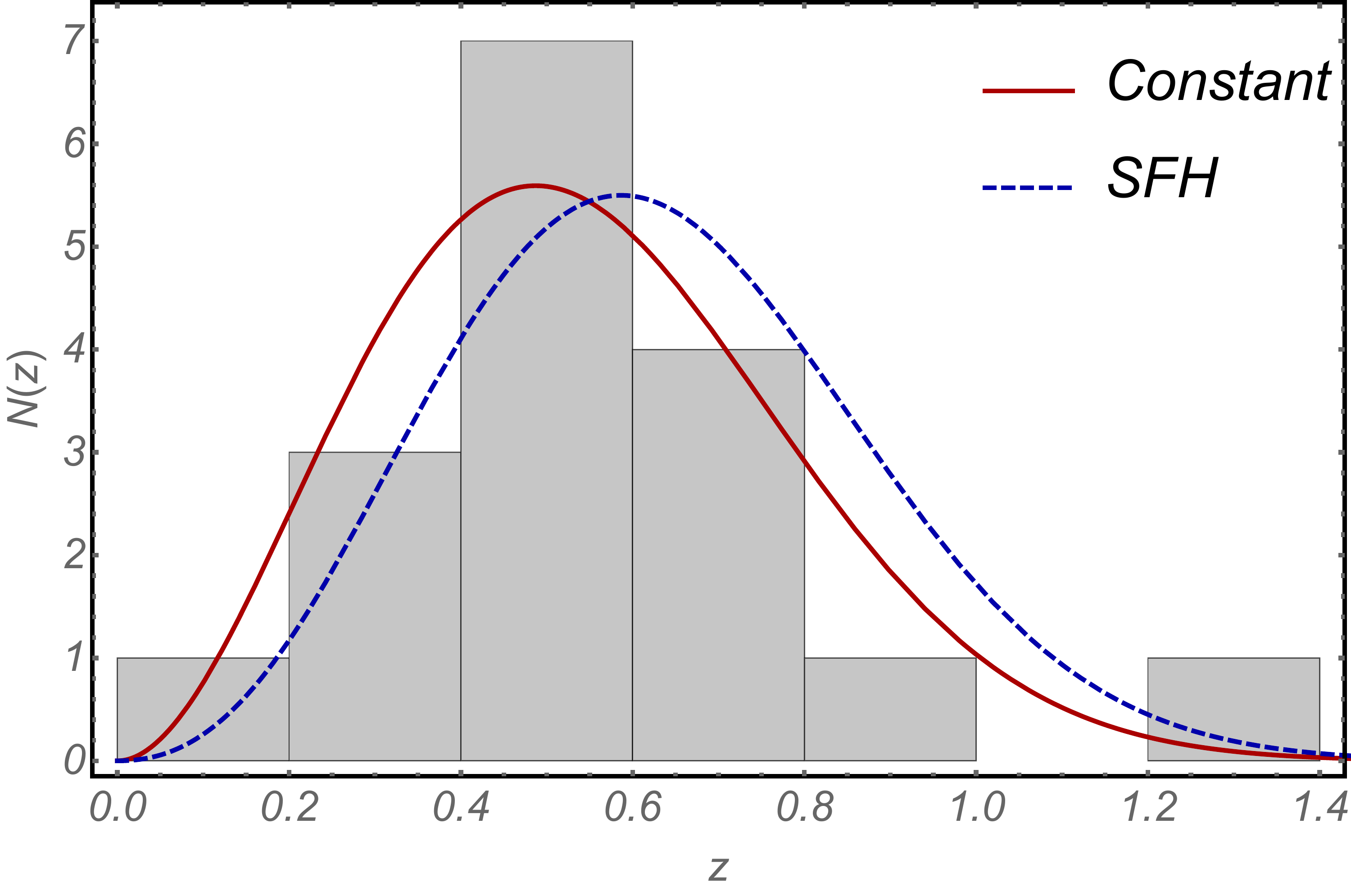}
	\caption{A histogram of the 17 FRBs observed to date, with inferred redshifts \cite{1601.03547}. 
	FRB redshift distributions are plotted assuming a constant comoving density 
	(solid-red), and following the star-formation history (dashed-blue), both with a cutoff 
	at $z_{\rm cut}=0.5$, and normalized to match the total number of detected events.}
	\label{fig:N(z)}
\end{figure}

In Figure~\ref{fig:N(z)} we plot a histogram of the estimated redshifts for the current 
FRB catalog \cite{1601.03547}, which is well fit by the two FRB distribution functions 
above, if a cutoff of $z_{\rm cut}=0.5$ is chosen.

To estimate the total number of FRBs observable in the near future, we consider an 
experiment like CHIME \cite{1406.2288}. In Ref.~\cite{1602.07292} it was estimated that CHIME will detect 
$\sim730-15000$ FRBs per year, and so we will take a fiducial, albeit optimistic, value of $N_{\rm FRB} = 10^4$ 
bursts per year. 

Interchannel dispersion broadens the FRB pulse arrival time to
\be
\delta t_{\rm DM} = 0.3 \, {\rm ms} \times \dfrac{\rm DM}{800 \,\rm pc\, cm^{-3}}\, \dfrac{\Delta\nu}{24\rm \,kHz}\left(\dfrac{800 \, \rm MHz}{\nu}\right)^3,
\ee
where $\nu$ is the frequency,
$\Delta \nu$ is the bandwidth, which will be 24 MHz, or smaller, for transient studies with CHIME \cite{1406.2288,Kiyo},
and DM is the dispersion measure, given by the integrated column density of electrons \cite{1512.02738,1506.01704}.
The total pulse width of a FRB will have a contribution from its (unknown) intrinsic pulse profile,
as well as scattering with the intergalactic medium \cite{Koay:2014gsa,1601.05410}, and
the lensing time delay has to be bigger than its total width to be easily detectable.
To account for this, we will require a lensing time delay longer than $\overline{\Delta t}=1$ 
ms as our baseline case. 
FRBs might have a distribution of intrinsic widths; wider bursts would give rise to 
more pessimistic results, whereas narrower FRBs might produce more optimistic ones.
We will therefore show results for $\overline{\Delta t}=0.3$ ms and $\overline{\Delta t}=3$ ms, as well.

Given how little is known about the luminosity function of FRBs \cite{1602.07292,1604.03909},
we will not attempt to model the lensing magnifications $\mu_+$ and $\mu_-$ observable at each source redshift.
Instead, we will simply require a constant flux ratio $\overline{R_f}=5$ as a threshold, since this will make the echoed image detectable.

\begin{figure}[htbp!]
	\includegraphics[width=0.45\textwidth]{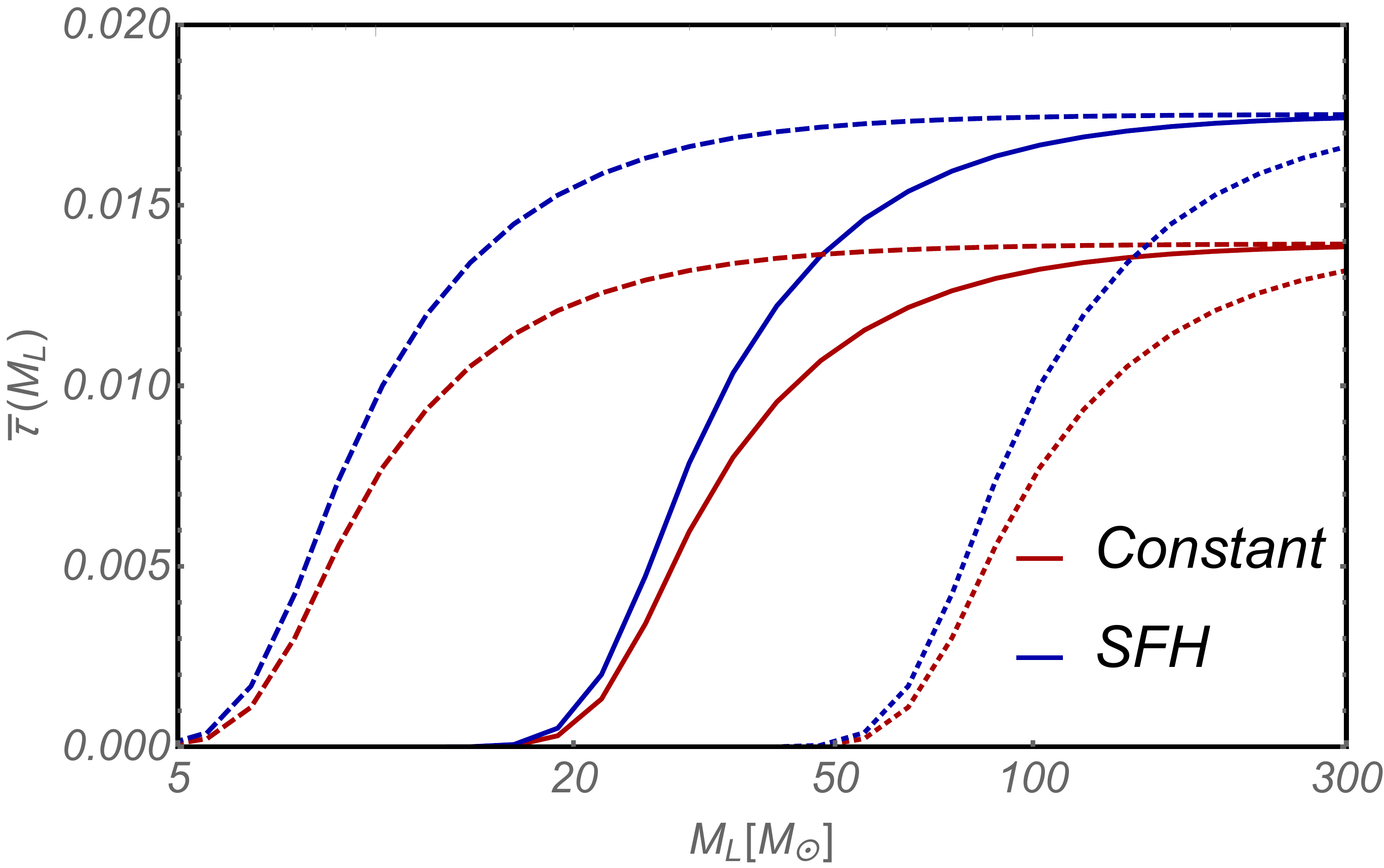}
	\caption{Integrated optical depth, with weightings corresponding 
	to a population	of FRBs with constant comoving density (red curves) and following 
	the SFH (blue curves), both with a cutoff at  $z_{\rm cut}=0.5$. 
	In dashed, solid, and dotted lines we require a time delay $\Delta t>$ 0.3, 
	1, and 3 ms, respectively. In all cases, $\fDM=1$.}
	\label{fig:taubar}
\end{figure}

Now, given a distribution function $N(z)$ for FRBs, we can calculate their integrated optical 
depth $\bar \tau(M_L)$, due to MACHOs of mass $M_L$, as
\be
\bar \tau (M_L) = \int dz\, \tau(z,M_L) N(z).
\ee
We show this quantity in Figure \ref{fig:taubar} for the same two 
distribution functions discussed above. 
It is clear that the distribution mimicking the SFH produces a higher optical depth, 
due to the higher redshift of most sources. 

We can finally forecast the number $N_{\rm lensed}$ of lensed FRBs that a fraction $\fDM$ 
of dark matter, in the form of point lenses of different masses, will yield. 
In all cases we are in the optically thin 
regime, where the probability to be lensed is just $P_{\rm lens} = 
1-e^{-\bar \tau}\approx \bar \tau$. Thus, if we observe a number 
$N_{\rm FRB}$ of FRBs, $\bar \tau  N_{\rm FRB}$ 
of them should be lensed. 
Notice that, even if all the dark matter was composed of compact objects of 
a single mass $M_L$, the lensing time delays induced on FRBs would not have a
unique value, due to the different impact parameters and redshifts of the lenses.

\begin{figure}[hb!]
	\includegraphics[width=0.48\textwidth]{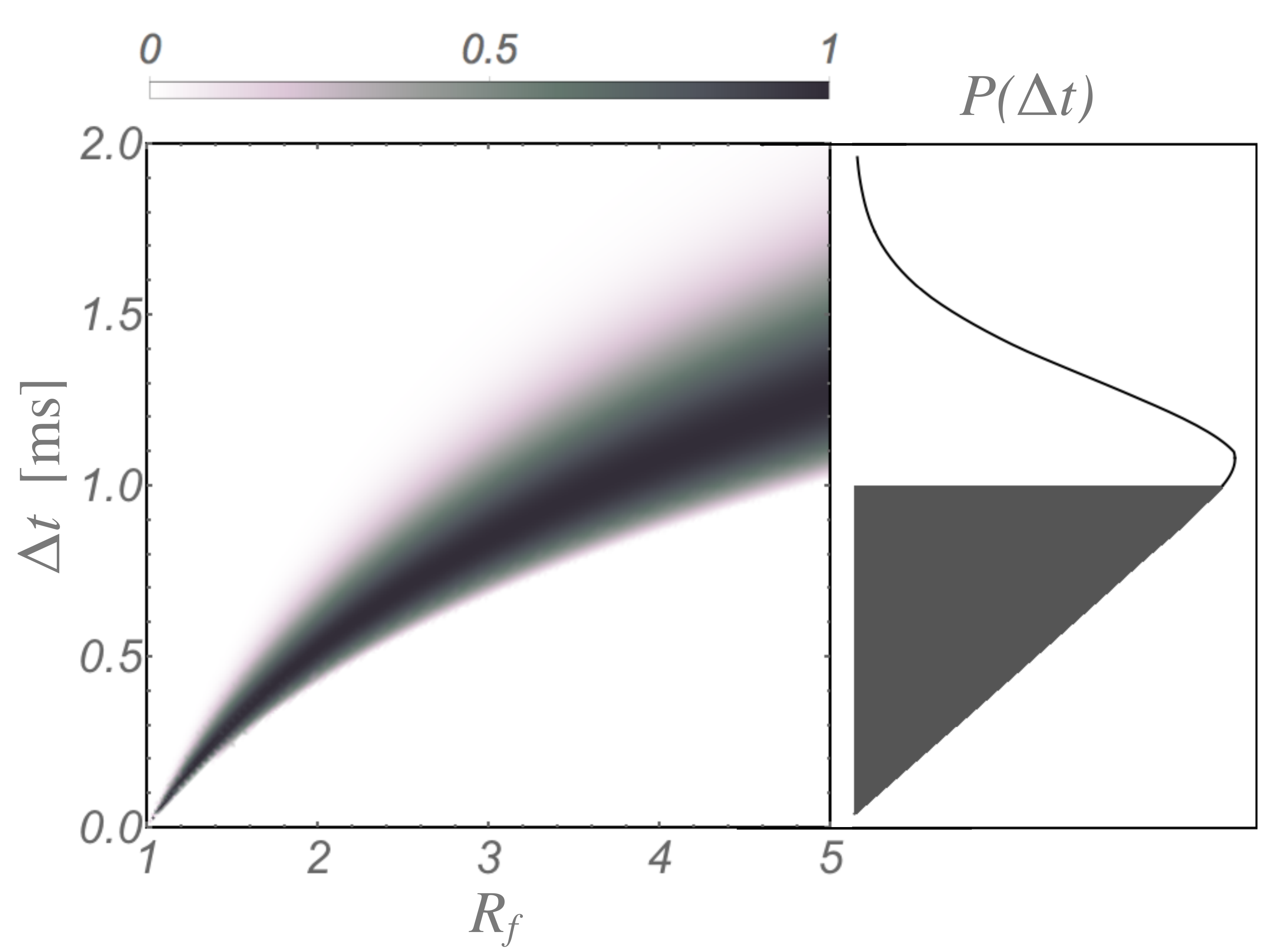}
	\caption{Joint probability distribution for the flux ratio $R_f$ and time delay $\Delta t$ between the two peaks of a FRB lensed by a 30 $M_\odot$ MACHO.  On the right, we marginalize over $R_f$, and show the probability to find a time delay $\Delta t$.  The shaded region corresponds to time delays smaller than 1 ms, too short to be detectable.
}
	\label{fig:PDFt}
\end{figure}

In Fig.~\ref{fig:PDFt} we show the joint probability distribution 
function (PDF) for a time delay $\Delta t$ and a flux ratio $R_f$, assuming a 30 $M_\odot$ lens.
This PDF has been calculated by convolving Eqs.~\eqref{eq:timedelay} and \eqref{eq:Rf},
assuming a flat distribution in impact parameters squared up to $y_{\rm max}^2$, with a population of FRBs following $N_{\rm const}(z)$,
and shows a clear correlation between the lensing time delays and the flux ratios.
We also show the probability $P(\Delta t)$
to find a time delay $\Delta t$ between the two events, calculated by marginalizing the PDF over $R_f<5$. 
This time-delay distribution
would be broadened further if the MACHOs had some range of masses instead of a single $M_L$.

Considering the most conservative case of a 
constant-density distribution of FRBs, with a cutoff at redshift $z_{\rm cut}=0.5$ as 
discussed above, and $10^4$
total detected FRBs, corresponding to one year of observation with CHIME,
we will see a number $N_{\rm lensed}= 13$ of lensed bursts with a time delay longer
than 1 ms, if all the dark 
matter is in the form of 20$\,\Msun$ MACHOs.
If, however, the dark matter is made of 30$\,\Msun$ PBHs, as suggested in Ref.~\cite{1603.00464}, 
the number of lensed events that will be detected is $N_{\rm lensed}= 60$. For all MACHO masses
larger than $M_L=100\,\Msun$ the number of lensed events is simply $N_{\rm lensed}= 130$.
Here we have required a flux ratio smaller than $\overline {R_f}=5$ to observe both bursts,
although high MACHO masses produce time delays much in excess of the threshold 
values of $\overline{\Delta t}$, so the cross section annulus in Eq.~\eqref{eq:tau} becomes a circle,
and $N_{\rm lensed}$ scales roughly linearly with $\overline{R_f}$.

\begin{figure}[htbp!]
	\includegraphics[width=0.48\textwidth]{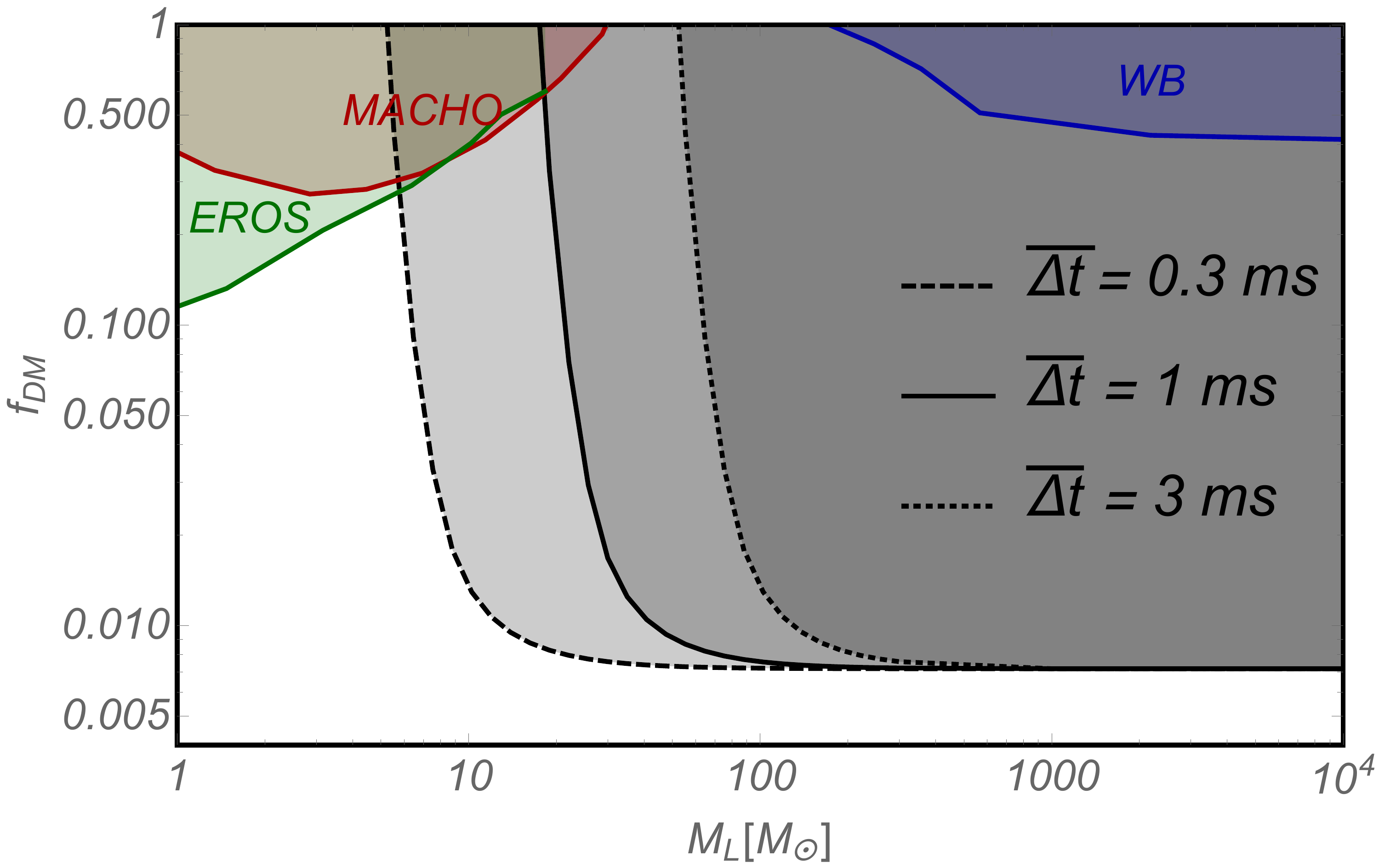}
	\caption{Fraction $\fDM$ of dark matter allowed in the form of point lenses of 
		mass $M_L$, if no events out of $N_{\rm FRB}=10^4$ are lensed, 
		where the FRBs have a constant comoving density with a cutoff at  $z_{\rm cut}=0.5$. 
		In dashed, solid, and dotted black we show our constraints when we require a time 
		delay $\Delta t>$ 0.3, 1, and 3 ms, respectively. In red we show the current 
		constraints from the MACHO Collaboration, in green the ones from the EROS 
		Collaboration, and in blue the constraints from galactic wide binaries.}
	\label{fig:fDM}
\end{figure}

We can also determine the smallest fraction $f_{\rm DM}$ that will produce at least one lensed event in a survey with $10^4$ FRBs. Fig.~\ref{fig:fDM} shows the regions of the $f_{\rm DM}$-$M_L$ parameter space that give rise to at least one such event for lensing time delays longer than 0.3, 1, and 3 ms.  
We also show the current constraints
to $\fDM$ from the EROS Collaboration \cite{astro-ph/0607207},
the MACHO Collaboration \cite{astro-ph/0011506}, 
and wide-binary disruption \cite{0903.1644}.

From Figure \ref{fig:fDM} we see that, if none of the $10^4$ upcoming FRBs is lensed, the amount of dark matter in MACHOs 
will be constrained to $\fDM< 0.8\%$ above a cutoff mass of $\sim 100\,\Msun$, 
under the assumption that the smallest time delay detectable is 1 ms. 
This will thus place more stringent constraints over this mass range than those 
coming from wide-binary disruption \cite{0903.1644}, by more than an order of magnitude.

For masses in the $20-100\,\Msun$ window, outside the reach of the Galactic-lensing 
surveys \cite{astro-ph/0011506}, a dark-matter fraction of $\fDM\sim\,8\%$, at the 
lower-mass end of this range, and $0.8\%$ at the higher-mass end, would suffice to detect 
one lensed FRB, if $10^4$ FRBs are observed with a time resolution of 1 ms.
As the number of lensed events scales trivially with 
$N_{\rm FRB}/10^4$, 
even a smaller number of $\sim 10^3$ FRBs per year should suffice to detect $\sim 1-10$ lensed
FRBs in the first year of operation of CHIME, if MACHOs in this window made up the dark matter. 
This conservative number will still
allow us to place constraints on $\fDM$, comparable in magnitude to all 
current surveys ($\fDM\leq 10\%$), but over the whole mass range $M_L>20\,\Msun$, if no lensed events are observed.
Interestingly, even with a time resolution of 3 ms one would detect at least one lensed event,
if MACHOs of mass $M_L\gtrsim\,50\,\Msun$ were the main component of dark matter.

FRBs might suffer intrinsic repetition. For example, the event FRB 121102
has been observed repeating as quickly as over minutes \cite{1603.00581}. 
Lensing by a MACHO of mass $M_L\sim 10^5 \Msun$ creates a time delay also on the scale of minutes, which then sets
a natural ceiling to the MACHO mass that can be unequivocally probed with lensing of FRBs.
In general, the correlation between time delays and flux ratios of the bursts, as shown in Fig.~\ref{fig:PDFt}, will be 
of invaluable help to statistically determine whether repetition of FRBs is caused by microlensing.

Throughout this work we have assumed that an upcoming CHIME-like experiment will detect events up to a cutoff 
redshift of $z_{\rm cut}=0.5$, as this fits the current FRB data. We can also calculate constraints 
for an increased cutoff redshift, e.g., $z_{\rm cut}=0.7$, representing a more optimistic 
redshift distribution.
In that case, for $\overline{\Delta t}=1$ ms and $\overline{R_f}=5$, we expect $N_{\rm lensed}=35$ lensed events out of $10^4$ if dark matter is made
of MACHOs of mass $M_L=20\,\Msun$,  $N_{\rm lensed}=110$ if this mass is $M_L=30\,\Msun$, and 
$N_{\rm lensed}\gtrsim200$ for masses higher than $75\,\Msun$. Were none of these $10^4$ FRBs
to show lensing, however, we could constrain $\fDM$ at 30 $\Msun$ to be smaller than $0.9\%$ (or  
$0.5\%$ for $\overline{\Delta t}=0.3$ ms, where this last number would apply to larger masses, 
and smaller $\overline{\Delta t}$). The increase in the lensing optical depth, due to 
the higher redshift of the events, leads to either more FRBs being lensed, or better constraints on $\fDM$.

It has been argued that we
could be preferentially observing strongly-lensed FRBs \cite{astro-ph/9511150,Fialkov:2016fjb}. 
If this is the case, most observed FRBs will be lensed by intervening objects, such as galactic 
halos, on their way to Earth. This would create a double image with a time delay on the order 
of weeks \cite{1403.7873}. More importantly, when crossing those galactic halos the probability to be microlensed by a MACHO is 
close to unity, which would help detect more microlensed FRBs, or improve our constraints on $\fDM$.

Note that, due to our requirement that they behave as point lenses, MACHOs 
need to be smaller than their Einstein radii. This constrains the size of a MACHO
of mass $M_L$ to be more compact than $\sim 0.1$ pc $\times \sqrt{M_L/30\,\Msun}$.

An effect similar to femto- or nanolensing of 
gamma-ray bursts could be observed in FRBs \cite{Gould,Zheng:2014rpa}, albeit, given the relatively low frequency 
($\nu \sim$ GHz) of FRBs, one could probe lenses only with masses higher than 
$M_L\sim 10^{-5}\Msun$, since lower masses would create a time delay smaller than $1/\nu$ and not cause interference. 
An experiment with bandwidth $\Delta\nu \sim 20$ kHz could probe a maximum mass  $M_L\sim 0.1\,\Msun$ with nanolensing 
(higher masses would cause time delays longer than $1/\Delta \nu$ and interfere within each bandwidth). 
The unknown FRB frequency spectrum poses a challenge 
to modeling this effect, so it is left for future work.

Among the FRBs found to date, there is one particular event, FRB 121002, which has 
been observed with a double peak delayed by 5.1 ms \cite{1511.07746}. This delay 
could have been caused by a MACHO lens of mass $M_L \gtrsim 200 \Msun$. The second image 
of FRB 121002 appears brighter, however, which contradicts the usual lensing 
prediction. 
In Ref.~\cite{us} we will assess how likely it is that this delay 
is due to lensing and study further cosmological applications of lensing of FRBs.

In conclusion, upcoming interferometers will open up the radio sky, which will allow 
us to detect FRBs at a staggering pace. By studying whether these FRBs are doubly peaked
we can conclude if they have been microlensed or not. Given the existing constraints, compact objects 
(MACHOs) are allowed to make up a large fraction of dark matter in our Universe (and even all of it
in the mass window between 20 and 100 $\Msun$). We will be able to 
detect from tens to hundreds of lensed FRBs if the dark matter is indeed composed of these MACHOs.
Alternatively, if no FRBs are microlensed, we will place 
the strongest constraints yet on the fraction of dark matter in the form of compact objects.

The authors thank Graeme Addison and especially Kiyoshi Masui for very valuable conversations. 
This work was supported at JHU by NSF Grant No. 0244990, NASA NNX15AB18G, the 
John Templeton Foundation, and the Simons Foundation. L.D. is supported by NASA through Einstein Postdoctoral Fellowship Grant No. PF5-160135 awarded by the Chandra X-ray Center, which is operated by the Smithsonian Astrophysical Observatory for NASA under Contract No. NAS8-03060.



\end{document}